\documentclass[conference]{IEEEtran}
\IEEEoverridecommandlockouts
\usepackage{cite}
\usepackage{amsmath,amssymb,amsfonts}
\usepackage{algorithmic}
\usepackage{graphicx}
\usepackage{textcomp}
\usepackage{xcolor}

\def\BibTeX{{\rm B\kern-.05em{\sc i\kern-.025em b}\kern-.08em
    T\kern-.1667em\lower.7ex\hbox{E}\kern-.125emX}}
\begin{document}

\title{VFNet: A Convolutional Architecture for Accent Classification}

\author{Asad Ahmed, Pratham Tangri, Anirban Panda, Dhruv Ramani, Samarjit Karmakar 
\\
Nevr\'onas
\\
National Institute of Technology, Warangal, India\\
{\tt\ ahmed.asad19@gmail.com, prathamtangri2015@gmail.com, anirban.panda.ndp@gmail.com} \\
\tt\ dhruvramani98@gmail.com, karmakar.samarjit@gmail.com}

\maketitle

\begin{abstract}
 Understanding accent is an issue which can derail any human-machine interaction. Accent classification makes this task easier by identifying the accent being spoken by a person so that the correct words being spoken can be identified by further processing, since same noises can mean entirely different words in different accents of the same language. In this paper, we present VFNet (Variable Filter Net), a convolutional neural network (CNN) based architecture which captures a hierarchy of features to beat the previous benchmarks of accent classification, through a novel and elegant technique of applying variable filter sizes along the frequency band of the audio utterances. 
\end{abstract}

\section{Introduction}

Accent refers to the distinctive way of speaking a particular language by a person. Pronunciation of different syllables, sometimes even usage of different words change in different accents. Often, people from the same region share the same accent, which eventually becomes an important part of the culture itself. There are three basic accentual methods: stress, tone, and length. But all three types of accentual systems aren't found together in a single language. 

English is one of the languages with the highest number of speakers in the world, and this has given rise to a plethora of accents from different places all around the world, such as American, British, Indian, Korean, Italian, etc. It is also the primary language for internet communication. Also, due to the extended work already done regarding accents in the language, we know neural network models can work with English accents comfortably. Thus, it makes it an ideal language to choose for an accent classification task. 

With the increase in the use of smart appliances interacting with the user based on speech, it has become increasingly important for the system to understand what a person is speaking correctly and accurately. Different accents for the same language can mean completely different commands even with similar audio characteristics. Unless the system involved can understand properly what the user is conveying, any interactive technology of this sort is incomplete. We can divide this problem into two parts: Finding the accent the speaker is speaking, and then finding what the speaker was saying. We here propose a new solution for the first issue. By knowing the accent of the speaker using the proposed architecture, we can enhance the quality of engagement with various speech-based virtual assistants and agents. 

In this paper, we propose an architecture trained on the audio spectrograms of raw audio utterances from the Speech Accent Archive~\cite{accent}.

\begin{figure*}
\centering
\includegraphics[height=60mm, width=130mm]{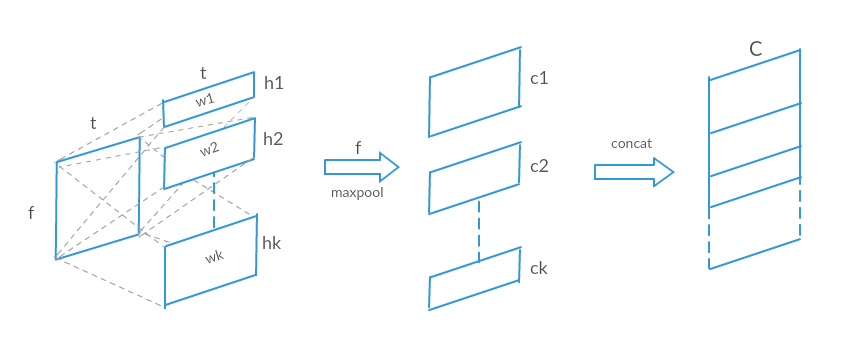}
\caption{A framework for accent classification which applies convolution with multiple variable filter sizes followed by maxpool and concatenation of filtered outputs. The signal is pre-processed by applying Short Time Fourier Transform (STFT) on the raw input audio to generate an audio spectrogram. This spectrogram is passed through the classification network to classify the signal. }
\label{figure1}
\end{figure*}

\section{Related Work}

Accent for a speech signal is different from other high level features such as phonemes or emotions. Thus, a limited number of works dealing with accent of a speech have been found. The task of Accent Classification was initially tackled by classical sequential models, such as Gaussian Mixture Models (GMMs) and Hidden Markov Models (HMMs). An architecture of parallel ergodic nets with context-dependent HMM units was utilized by Teixeira et al~\cite{teixeira}. While Deshpande et al~\cite{deshpande} used GMMs based on formant frequency features to discriminate between standard American English and Indian accented English.

Although Artifical Neural Networks (ANN), including Deep Neural Networks (DNNs) and Recurrent Neural Networks (RNNs), have been used in several areas of machine learning research since a long time with stellar results, in the field of accent classification there are only a few studies which evaluated their applicability. The paper of Jiao et al~\cite{rnndnn:16} is one of them, which presented an accent identification system by combining DNNs and RNNs trained on long-term and short-term features respectively. They tried to identify the native languages of non-native English speakers from eleven countries and proposed a combination of long-term and short-term training based on the observations that accent differences are mostly due to prosodic and articulation characteristics. Their research showed that the use of Neural Networks (NNs) together with Support Vector Machines (SVMs) outperformed NN alone and SVM alone.

In our paper, we have worked on classifying Arabic, Mandarin and native English accents accurately using a convolutional architecture.

\section{Problem Definition and Formulation}

The task of accent classification can be defined as follows:

For an audio utterance $X$, our task is to determine the mapping $F$, parameterized by $\lambda$, from the input space of audio utterances to the output space of accent labels. Hence, $F_{\lambda}(X)=l$, where $l$ is the accent of the audio speech utterance of $X$.

\section{Our Proposed Architecture}
To capture the accent of a speech signal, we propose a convolutional architecture with variable filter size. The architecture consists of a convolutional layer, processed with different filters with variable sizes, followed by maxpool and concatenation of various filtered outputs, which is passed through a fully connected layer to get the accent of the speech signal. 

\subsection{Pre-processing} 
The existing accents of speech signal differ from each other in terms of frequency contour or rhythmic features. However, by analyzing the transformed features of the signal, it was found that the relative stress in particular frequency helps a lot in the classification of accent. Thus, the raw audio was processed with Short Time Fourier transform (STFT), by which the time domain signal was converted into the frequency domain. After that, the frequency domain signal was converted into magnitude-spectral domain by taking the magnitude of the result of STFT. The spectrum of the signal, in the magnitude-spectral domain, was obtained by taking a log of magnitude with time as the horizontal axis and frequency as the vertical axis.  

Mathematically, let $x(t)$ be the input raw audio signal. The spectrogram for the signal $x(t)$ for a window function $w$ is given by, 
\begin{equation} 
    Spec\{x(t), w\}=log_{e}(|STFT\{x(t), w\}|) 
\end{equation} 
where, the $STFT$ function is given by,
\begin{equation}
    STFT\{x(t), w\}= \int_{-\infty}^{+\infty}x(t)w(t-\tau)e^{-j\omega t}dt
\end{equation}

\begin{figure}
\centering
\includegraphics[height=40mm, width=40mm]{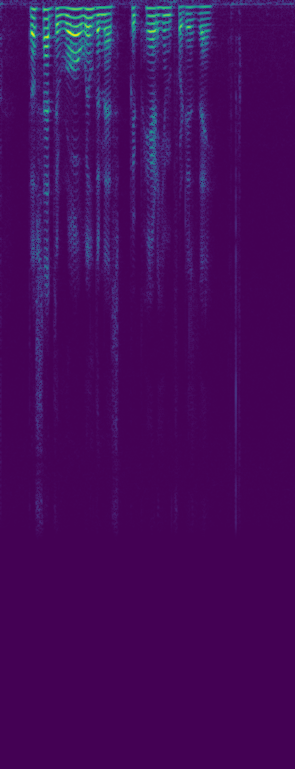}
\caption{The sprectrogram representation of a raw audio signal after processing with STFT and bringing it to the magnitude-spectral domain.}
\label{trainacc}
\end{figure}

\subsection{Classifier Network}

Let the spectrogram be represented by $X \in \mathcal{R}^{f \times t}$, where $f$ represents the frequency axis and $t$ represents the time axis. We apply convolution with multiple variable filter sizes $w_i \in \mathcal{R}^{h_i \times t}$, where $h_i$ represents the height of the $i^{th}$ filter, $i \in [1, k]$.

\begin{equation}
    c_i = f(w_i * X + b)
\end{equation}

Here, $b \in \mathcal{R}$ represents bias and $f$ represents the ReLU non-linearity function. Multiple such operations are performed with variable windows sizes and concatenated into one single feature vector $c$, which is passed through fully connected layers for classification.
\begin{equation}
    c = c_1 \oplus c_2 \oplus ... \oplus c_k
\end{equation}
The idea is to capture accent relevant features, which could be a mixture of low, medium and high-level features. We take variable filter sizes on a fixed time segment to capture the varying frequency axis and hence, capturing a hierarchy of accent relevant features. The concatenated vector $c$ represents the whole feature vector of accent relevant features of the audio spectrogram $X$.

\section{Experiments}

\begin{figure}
\centering
\includegraphics[height=50mm, width=70mm]{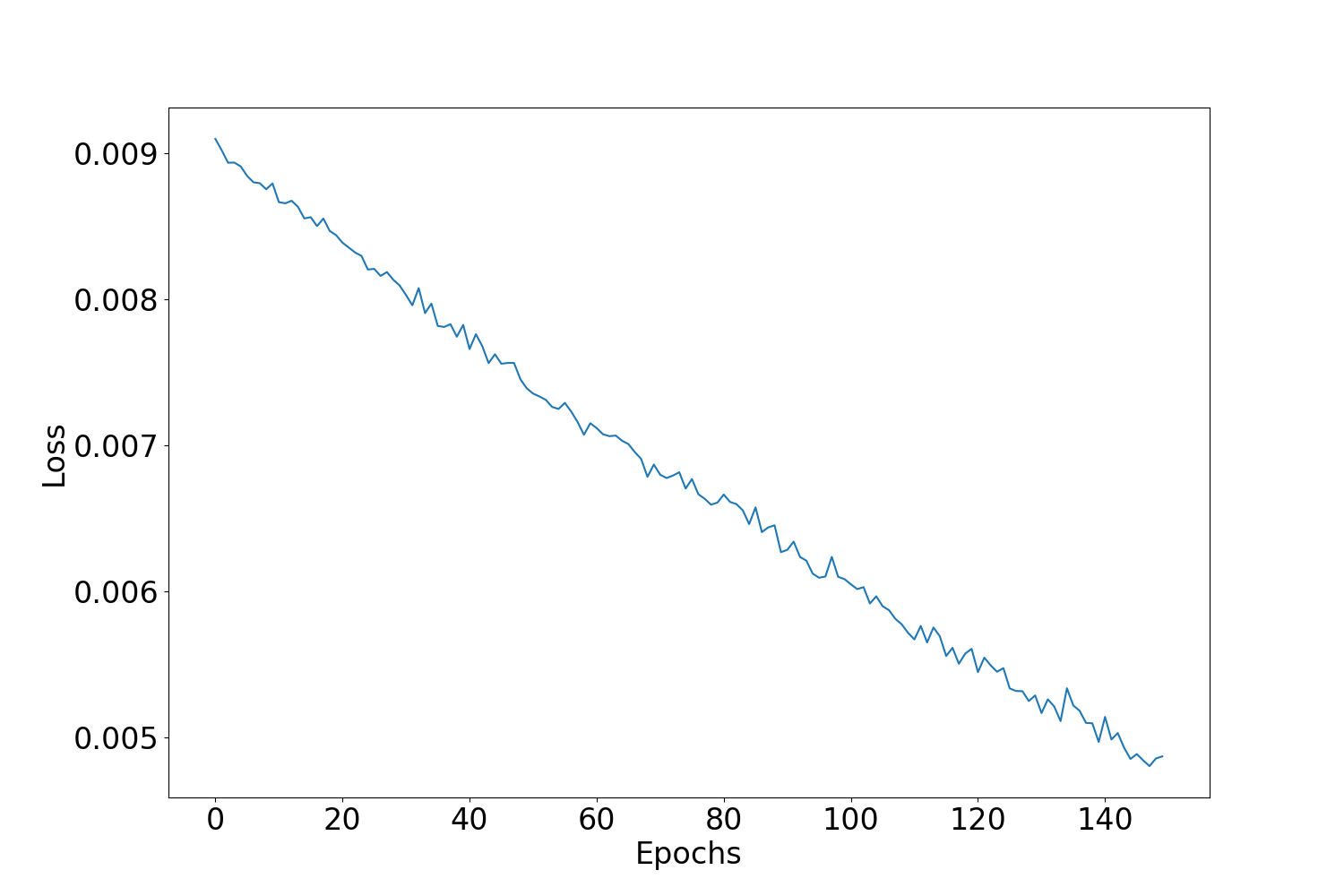}
\caption{
    The plot of training loss over a few epochs, on the training set, which clearly shows a decreasing nature as the number of epochs increases.}
\label{trainloss}
\end{figure}

\begin{figure}
\centering
\includegraphics[height=50mm, width=70mm]{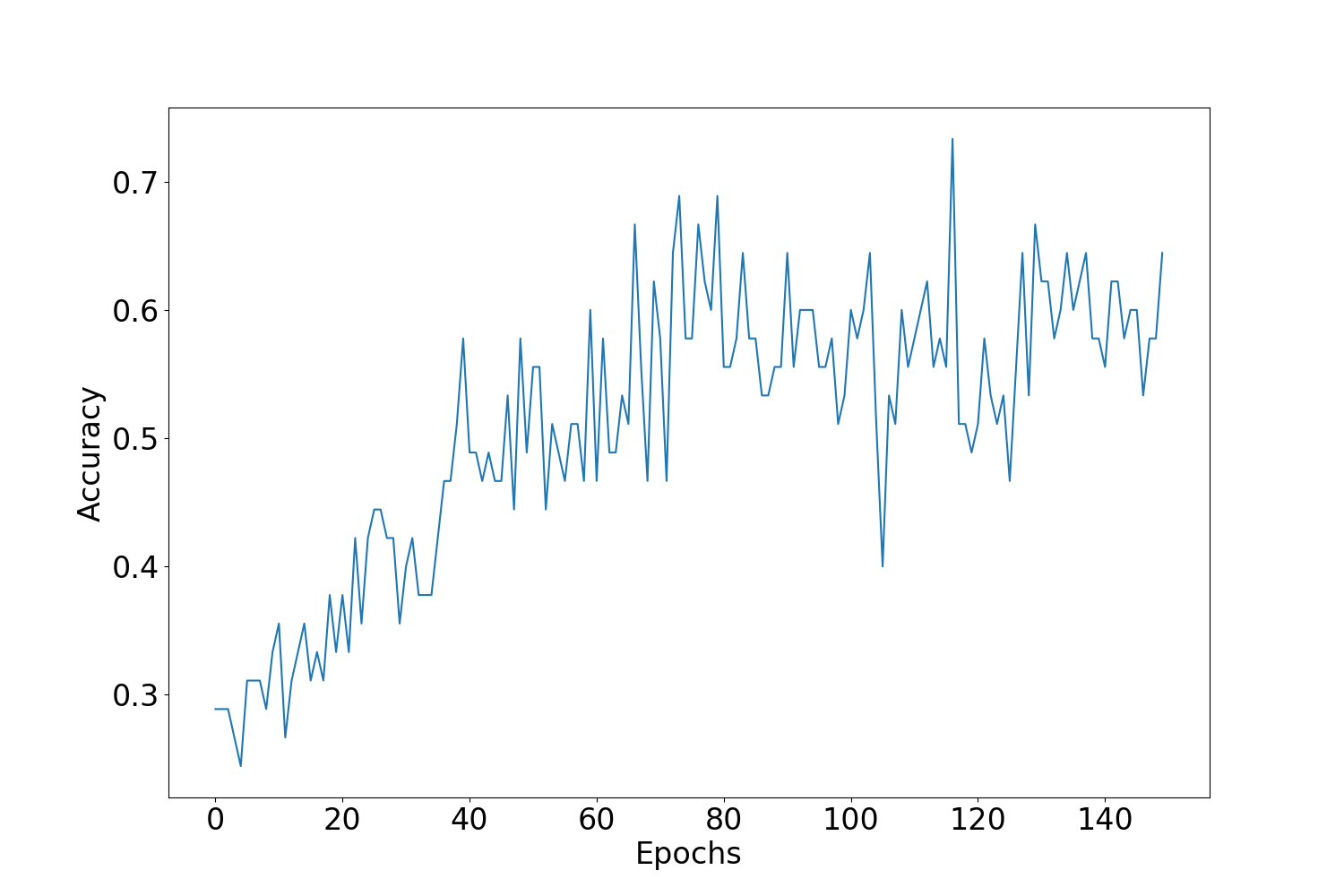}
\caption{The plot of testing accuracy over a few epochs, on the testing set, which ensures the convergence of the model after several epochs of training.}
\label{trainacc}
\end{figure}
The accent classifier network is trained on preprocessed content samples taken from the Speech Accent Archive~\cite{accent}, a repository of audio files which consist of a sentence spoken by over 2000 speakers in over 100 accents. The dataset provides speech signals and their corresponding accent labels. The corpus contains clean speech from 109 speakers with different accents which read out the same sentence in the English language. Out of all the audios present, we used some 74 female speakers with a same accent for training our networks. We downsampled the audio to 16 kHz for our convenience.

The raw audio signal is converted to a spectrogram by applying Short Term Fourier Transform (STFT) which brings the speech utterance to the frequency domain. This is then followed by an application of $log$ to the magnitude of the signal. The signal is now in the form of the magnitude-spectral domain. For better convergence and generalization of the networks, the spectrogram is split with columns of size $120$, and all of them are labelled with the corresponding targets and saved. These are then randomly sampled from the dataset, and hence create a segmented spectrogram.

The loss has been minimized using Adam~\cite{adam:15} as an optimizer, with learning rate as 0.001. To compare our model with the other existing models, we trained using Alexnet~\cite{alex}, Resnet~\cite{resnet}, and the model explaned in ~\cite{rnndnn:16}. The models have been implemented in the PyTorch deep learning framework and trained on a single Nvidia GTX 1070 Ti GPU. 

\section{Results and Discussion}

\begin{table}
\caption{Comparison of accent classifier models}
\begin{center}
 \begin{tabular}{||c | c ||} 
 \hline
 Model & Accuracy  \\ [0.5ex] 
 \hline
 CNN (AlexNet based)~\cite{alex} & 44.37 \\
 \hline
 CNN (ResNet based)~\cite{resnet} & 47.64 \\
 \hline
 RNN+DNN~\cite{rnndnn:16} & 59.88  \\ 
 \hline
 Our Model (VFNet) & \textbf{70.33}  \\ [1ex] 
 \hline
\end{tabular}
\end{center}
\end{table}

\begin{table}
\caption{Confusion Matrix}
\begin{center}
 \begin{tabular}{||c | c | c | c||}
 \hline
  & ENG & ARA & MAN  \\ [0.5ex] 
 \hline
 ENG & \textbf{90}  & 6 & 44 \\ 
 \hline
 ARA & 0  & \textbf{71} & 6 \\ 
 \hline
 MAN & 10 & 23 & \textbf{50}  \\ [1ex] 
 \hline
\end{tabular}
\end{center}
\end{table}

The trained accent classifier network was evaluated on several content audio clips. The network classified the audio clips to a particular accent label. The loss function decreases over a few epochs as shown in Figure~\ref{trainloss}. The accuracy of the model increases as the number of epochs increase. The accuracy of the model was compared with other various existing accent classifiers and is shown in Table 1.

It has been evident that our method of classifying accents performs increasingly better than the earlier models on similar approaches for accent classification. The confusion matrix is shown in Table 2. The probable reason for better performance of our model is that we captured a mixture of low, medium and high-level features to classify the accent. The variable filter sizes on a fixed time segment allowed us to capture the varying frequency axis and hence, capturing a hierarchy of accent relevant features.

The variable filter inspired model was able to achieve 90\% classification accuracy for native English accent, 71\% accuracy for Arabic accent and 50\% accuracy for Mandarin accent.

\section{Conclusion}
The accent of a speech signal is different from other high-level features of speech and is difficult to capture using simple convolutional networks. Hence, we tried to solve the task of accent classification using a new classifier network with variable filter sizes. In this way, specific accent features were found which depends on variable frequency, capturing a hierarchy of representations from low to high level. The overall accuracy of our approach outpaces the accuracy of other accent classifier based papers earlier.

\section{Acknowledgements}
This work is supported by Innovation Garage (IG), National Institute of Technology, Warangal.

\end{document}